\documentclass[conference]{IEEEtran}
\IEEEoverridecommandlockouts

\usepackage{cite}
\usepackage{amsmath,amssymb,amsfonts}
\usepackage{algorithmic}
\usepackage{graphicx}
\usepackage{textcomp}
\usepackage{xcolor}
\usepackage{booktabs}
\usepackage{multirow} 
\usepackage{adjustbox}

\usepackage{hyperref}
\newcommand{\model}{EventHunter} 

\def\BibTeX{{\rm B\kern-.05em{\sc i\kern-.025em b}\kern-.08em
    T\kern-.1667em\lower.7ex\hbox{E}\kern-.125emX}}
\begin{document}

\title{\model{}: Dynamic Clustering and Ranking of Security Events from Hacker Forum Discussions}

\author{
    \IEEEauthorblockN{
        Yasir ECH-CHAMMAKHY\IEEEauthorrefmark{1}\IEEEauthorrefmark{2},
        Anas Motii\IEEEauthorrefmark{1},
        Anass Rabii\IEEEauthorrefmark{2},
        Jaafar Chbili\IEEEauthorrefmark{3}
    }
    \IEEEauthorblockA{
        \IEEEauthorrefmark{1}College of Computing, Mohammed VI Polytechnic University (UM6P), Ben Guerir, Morocco \\
        \IEEEauthorrefmark{2}Deloitte Morocco Cyber Center, Casablanca, Morocco \\
        \IEEEauthorrefmark{3}Deloitte Conseil, Paris, France
    }
    \IEEEauthorblockA{
        Emails: \IEEEauthorrefmark{1}\{Yasir.ECH-CHAMMAKHY, Anas.MOTII\}@um6p.ma;
        \IEEEauthorrefmark{2}arabii@deloitte.fr;
        \IEEEauthorrefmark{3}jchbili@deloitte.fr
    }
}

\maketitle

\begin{abstract}
Hacker forums provide critical early warning signals for emerging cybersecurity threats, but extracting actionable intelligence from their unstructured and noisy content remains a significant challenge. This paper presents an unsupervised framework that automatically detects, clusters, and prioritizes security events discussed across hacker forum posts. Our approach leverages Transformer-based embeddings fine-tuned with contrastive learning to group related discussions into distinct security event clusters, identifying incidents like zero-day disclosures or malware releases without relying on predefined keywords. The framework incorporates a daily ranking mechanism that prioritizes identified events using quantifiable metrics reflecting timeliness, source credibility, information completeness, and relevance. Experimental evaluation on real-world hacker forum data demonstrates that our method effectively reduces noise and surfaces high-priority threats, enabling security analysts to mount proactive responses. By transforming disparate hacker forum discussions into structured, actionable intelligence, our work addresses fundamental challenges in automated threat detection and analysis.
\end{abstract}


\section{Introduction}
\label{sec:introduction}

The evolving landscape of cyber threats demands increasingly sophisticated detection and response capabilities~\cite{CyberMagazine2024, chen2024survey, Elouardi2024HybridCNNLLM}. Hacker forums have emerged as pivotal sources of cyber threat intelligence (CTI), facilitating the exchange of illicit tools, tactics, and sensitive information critical to executing advanced cyber attacks. These platforms have become significant marketplaces, generating substantial revenues, such as \$300 million annually from credit card fraud and \$864 million from DDoS-for-hire schemes~\cite{zhang2018idetector}. Therefore, continuous and effective monitoring of hacker forums is essential for proactive threat detection.

Researchers have explored various techniques to derive actionable intelligence from these platforms, including identifying influential hackers~\cite{amadou2024hc, amadou2024eurekha, samtani2017exploring}, mapping community structures~\cite{manatova2024understand}, extracting cybersecurity neologisms~\cite{li2021nedetector}, and classifying text to filter relevant discussions~\cite{deliu2018collecting, kadoguchi2019exploring, fang2019analyzing, moreno2023cream}. However, these approaches rarely focus on the automated detection and aggregation of security-related events as they unfold within forum discussions. Aligning with similar conceptualizations in security text analysis~\cite{tang2023trigger, cui2024tweezers}, we define a \textit{security event} in this context as a collection of related forum activities, such as discussions, announcements, transactions or information sharing that center on a specific cybersecurity occurrence. This includes, but is not limited to, the emergence of zero-day vulnerabilities, the distribution or sale of malware and exploits, the disclosure of breached data, or the offering of illicit attack services.

Identifying these events is critical because hacker forums often serve as crucial early warning systems, surfacing discussions related to significant security incidents well before they appear in conventional threat intelligence reports or public disclosures~\cite{paladini2024hackerforums, sapienza2017early}. Because threat actors directly leverage these platforms to initiate, propagate, or orchestrate attacks, monitoring these emerging events offers potentially vital lead time for defensive measures. However, this research gap persists partly because CTI efforts have often prioritized platforms like Twitter, which offer relatively centralized data streams compared to the fragmented ecosystem of numerous and disparate hacker forums~\cite{rahman2023what, cui2024tweezers, yagcioglu2019detecting}. Furthermore, monitoring these forums poses inherent challenges, including user anonymity, widespread jargon, rapid content velocity, and deliberate obfuscation tactics~\cite{paladini2024hackerforums}.

Despite their analytical value, the volume, noise and fragmented nature of discussions across numerous forums render manual monitoring and timely identification of critical events operationally infeasible~\cite{sans2023ctisurvey, Cybersixgill2024}. This challenge directly contributes to analyst overload and alert fatigue within Security Operations Centers (SOCs)\cite{sans2023ctisurvey, chen2024survey}. Even after initial filtering identifies potentially relevant forum posts, the large volume can exceed analysts' capacity for investigation. Therefore, analysts often lack the resources to thoroughly investigate all potentially true signals, not just filter false positives\cite{sadlek2025severity}. Addressing this matter requires moving beyond simple filtering to effective prioritization, allowing analysts to focus limited resources on the events posing the greatest potential risk. Therefore, our primary objective is to develop \model{}, an automated framework designed for both detecting emerging security events through clustering fragmented discussions and enabling systematic event ranking based on quantifiable metrics reflecting potential impact and importance. This approach aims to provide analysts with timely, actionable intelligence, facilitating focused investigation and proactive response. The main contributions of this paper are threefold:
\begin{enumerate}
    \item We propose \model{}, an unsupervised framework for dynamically detecting security events by clustering fragmented forum posts using a novel entity-aware contrastive text embedding model specifically designed for the linguistic characteristics of hacker forums.
    
    \item We introduce a systematic, data-driven event prioritization methodology that objectively ranks detected event clusters based on quantifiable metrics for timeliness, relevance, credibility, and completeness, providing analysts with actionable intelligence.
    
    \item We present a comprehensive empirical evaluation and measurement study using large-scale, real-world hacker forum data, demonstrating \model{}'s effectiveness in identifying coherent security events and prioritizing high-impact threats, supported by detailed performance metrics and case studies.
\end{enumerate}

The rest of this paper is organized as follows: Section~\ref{sec:related_work} discusses related work. Section~\ref{sec:methodology} details the \model{} framework and its components. Section~\ref{sec:experiments} presents the experimental setup and evaluation results. Section~\ref{sec:discussion} discusses the findings, limitations, and implications, including the identification of potential key actors. Finally, Section~\ref{sec:conclusion} concludes the paper.

\begin{figure*}[t!]
    \centering
    \includegraphics[width=.9\textwidth]{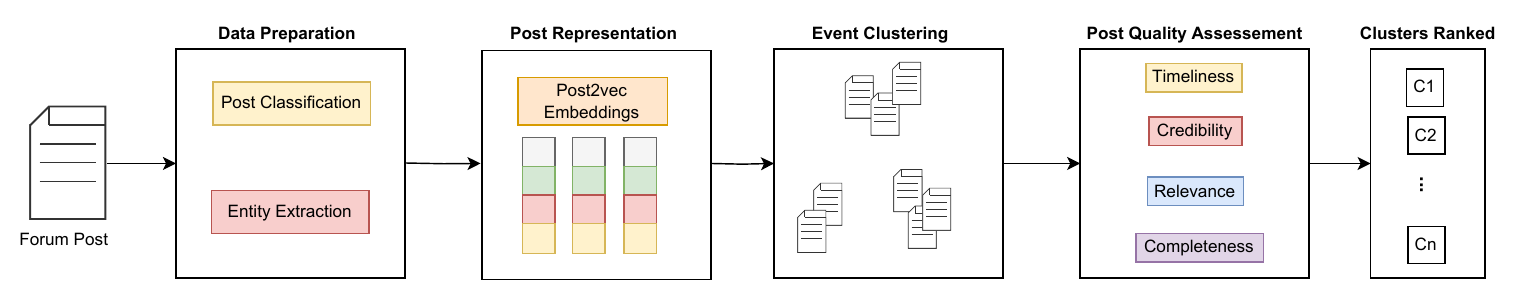}
    \caption{Overview of the \model{} framework: Forum posts are processed through data preparation (including classification and entity extraction) and transformed into dense, entity-aware post embeddings. These embeddings are then used to group related posts into event clusters. Finally, these clusters are assessed based on key quality metrics (Timeliness, Credibility, Relevance, Completeness) and ranked for analyst prioritization.}
    \label{fig:cti_workflow}
    \vspace{-5pt} 
\end{figure*}

\section{Related Work}
\label{sec:related_work}

Event detection is a central task in information retrieval and natural language processing, aimed at identifying and characterizing evolving events from data streams. This task has been explored across various domains, including natural disasters~\cite{Ashktorab2019Using, Hagras2017Towards}, public health surveillance~\cite{Parekh2024Event, Yousefinaghani2019Assessment}, and real-time news tracking~\cite{Alkhodair2020Detecting}. A core challenge in event detection lies in extracting semantically coherent occurrences from high-volume, unstructured data, while effectively managing constraints related to scale, velocity, and noise~\cite{Fedoryszak2019RealTime}.

In cybersecurity, identifying events early is fundamental for effective cyber-threat intelligence (CTI)~\cite{chen2024survey}. Online platforms, particularly social media like Twitter and underground forums, serve as critical CTI sources, often providing early indicators of vulnerabilities, threat actor activities, and illicit services~\cite{zhang2018idetector, sapienza2017early}.

Recent work has validated the importance of this area; notably, Paladini et al.~\cite{paladini2024hackerforums} conducted a large-scale longitudinal study demonstrating that forum discussions often precede official security reports. Their work provides crucial measurement-based validation for monitoring these platforms, answering ``\textit{are forums still early?}''. Our work addresses a different, complementary problem: we propose \model{}, an unsupervised framework designed not for historical analysis, but to answer ``\textit{what specific event on this forum should I investigate right now?}'', thereby shifting the focus to operational intelligence generation.

Early detection methods frequently used trigger-based techniques using predefined keyword sets to identify potentially relevant security posts~\cite{sceller2017sonar, shin2020cybersecurity}. While computationally efficient, such static approaches are not efficient in adversarial environments like hacker forums, where obfuscated language, neologisms, and evolving terminology limit their recall and robustness.

More recent approaches have advanced beyond simple keyword matching by incorporating semantic representations derived from contextual embeddings~\cite{yagcioglu2019detecting, cui2024tweezers, tang2023trigger}. This innovation enables better detection of novel or linguistically varied threats missed by keyword-based systems. However, many such methods were primarily developed or evaluated on relatively centralized platforms (e.g., Twitter) or more formal text sources. Their direct applicability is limited when dealing with the unique characteristics of the hacker forum ecosystem, which is marked by high noise levels, fragmented discussions across numerous disparate platforms, and rapid shifts in language and topics~\cite{paladini2024hackerforums}. Furthermore, much existing research has concentrated on supervised classification of individual posts, such as identifying security incident reports~\cite{yagcioglu2019detecting, fang2020detecting, khandpur2017crowdsourcing} or discussions about data breaches~\cite{fang2019analyzing, dong2018new}. While useful for initial filtering, classifying individual posts differs significantly from identifying broader thematic trends (topic modeling) or, as addressed in our work, aggregating fragmented posts into coherent, actionable \textit{events}. Our work tackles this more complex challenge: unsupervised clustering of semantically related posts into distinct events, a capability particularly crucial given the dynamic nature of threat landscapes where specific incidents unfold across multiple discussions over time.

While standard semantic embeddings capture general context, their precision for distinguishing specific security events can be reduced by the noise and ambiguity prevalent in forum data. Named Entities (NEs)—such as specific malware names, vulnerability identifiers (CVEs), threat actor handles, and targeted organizations—serve as crucial anchors in cybersecurity discussions, often defining the core elements of a security event~\cite{arazzi2023nlp, rahman2023what}. Leveraging NEs to link and group related posts offers a potentially more robust approach to event detection by focusing on these shared critical elements~\cite{komecoglu2024event}. For instance, discussions across multiple threads or forums referencing the same specific vulnerability identifier (e.g., CVE-2025-XXXX) likely pertain to the same underlying security event. \model{} directly incorporates this insight by employing an unsupervised, entity-aware contrastive learning approach tailored to generate embeddings sensitive to these key entities within noisy forum text.

A second significant limitation in the existing literature is the general absence of principled event prioritization mechanisms. While some studies perform post-clustering ranking~\cite{bose2019novel, cui2024tweezers}, they often rely on simple heuristics like cluster size or recency, which may correlate poorly with the actual impact or urgency of the detected event. High-severity events might initially generate only sparse discussion, while lower-relevance topics could produce substantial post volume. Effective CTI necessitates not only detection but also triage capabilities to help analysts prioritize events according to operational relevance and potential risk.

Current CTI quality frameworks emphasize dimensions such as timeliness, credibility, completeness, and relevance as essential for actionable intelligence~\cite{tundis2022feature, geras2024quality, zibak2022threat}. However, existing automated event detection and ranking models largely overlook the systematic integration of these factors. Our work addresses this gap through a dedicated prioritization module that evaluates discovered event clusters based on quantifiable metrics aligned with these four key dimensions. This methodology aims to shift CTI systems from purely detection-focused tools towards operationally valuable intelligence platforms that actively support analyst decision-making and resource allocation in time-sensitive SOC environments.

\section{Methodology: The \model{} Framework}
\label{sec:methodology}

This section presents the \model{} framework, which describes our unsupervised methodology to dynamically discover and prioritize security events within hacker forums. We describe the process of transforming fragmented forum posts into rich representations, clustering related discussions into coherent event timelines, and ranking these events for analyst attention.

\subsection{Problem Statement}
The distributed and dynamic nature of hacker forums presents distinct challenges for security event detection compared to traditional CTI sources. Although vendor reports typically consolidate comprehensive details within well-defined documents, forum discussions produce fragmented and continuously evolving information dispersed across numerous individual posts. This fragmentation complicates the detection and aggregation of security events, requiring a formal framework for tracking and connecting related content.

Formally, we approach this challenge by processing a continuous stream of hacker forum posts:
\[
\mathcal{P} = \{p_1, p_2, \ldots, p_n\},
\]

Our primary objectives are to:
\begin{enumerate}
    \item Derive \emph{security event clusters}:
    \[
    \mathcal{C} = \{c_1, c_2, \ldots, c_m\},
    \]
    where each cluster \( c_i \) groups forum posts \( p \in \mathcal{P} \) discussing a specific security incident.

    \item Generate a prioritized list of these events:
    \[
    \mathcal{R} = \text{sort}\left(\{(c_1, s_1), (c_2, s_2), \ldots, (c_m, s_m)\}\right),
    \]
    where \( s_i \) represents the calculated \textbf{Priority Score} for cluster \( c_i \), and the list \( \mathcal{R} \) is ordered by \( s_i \) to guide the analyst's attention. 
\end{enumerate}

Formally, this process can be expressed as:
\[
\mathcal{C} = f_{\text{cluster}}(\mathcal{P}),
\quad
\mathcal{R} = f_{\text{rank}}(\mathcal{C}),
\]
where \( f_{\text{cluster}} \) encapsulates the automated clustering mechanism that uses content similarity and contextual features, and \( f_{\text{rank}} \) quantifies the priority of each cluster based on metrics reflecting the timeliness, relevance, credibility, and completeness derived from the constituent posts.

\begin{figure*}[t]
    \centering
    \includegraphics[width=0.9\textwidth]{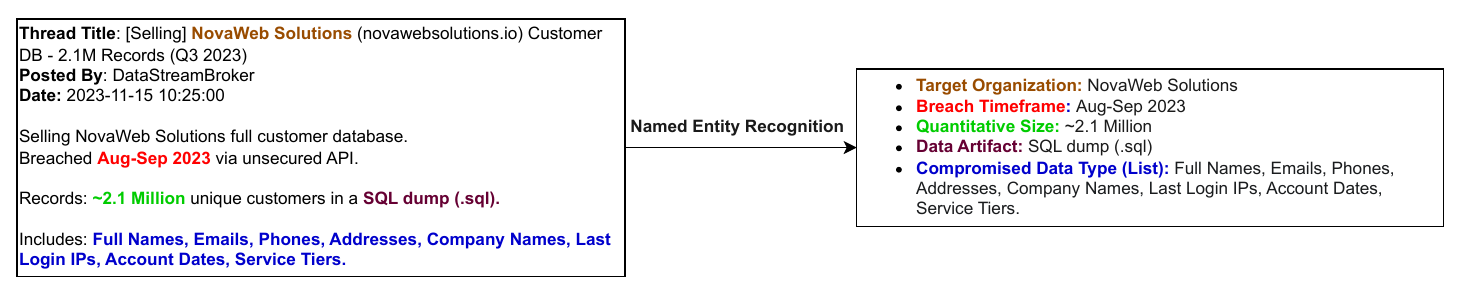}
    \caption{Illustration of CTI entity identification within a forum post announcing a data breach.
    }
    \label{fig:ner_example}
    \vspace{-5pt}
\end{figure*}

\subsection{System Overview}
\label{sec:system_overview}

As depicted in Figure~\ref{fig:cti_workflow}, the \model{} framework operates through five sequential stages:

\noindent\textbf{Data Acquisition.}
\label{subsec:data_acquisition_overview} 
We utilized the CrimeBB dataset~\cite{Pastrana2018CrimeBB}, encompassing over 91 million posts from 43 distinct cybercrime forums. The scale of this dataset provides a comprehensive foundation for observing global cybercriminal activities and discussions.

\noindent\textbf{Data Preparation.}
\label{subsec:data_preparation_overview}
Raw forum data is inherently noisy and unstructured. To prepare the acquired posts for effective representation and analysis, we perform essential preprocessing steps. Initially, post classification filters the data stream to retain relevant cybersecurity discussions and reduce noise~\cite{deliu2018collecting}. After that, Named Entity Recognition (NER) is applied to extract structured cybersecurity entities from these posts~\cite{chen2023enhancing, mouiche2025entity}. Both the classifications and the extracted entities serve as crucial inputs, significantly informing the downstream clustering process and the final event prioritization ranking.

\noindent\textbf{Post Representation.}
\label{subsec:post_representation_overview} 
Each incoming forum post \( p \) is transformed into a meaningful vector representation. As described in Section~\ref{subsec:post_representation}, we generate a dense, entity-aware embedding \( \mathbf{h}_p \) for each post using a fine-tuned Transformer model trained with a contrastive learning objective to capture semantic and entity-based relatedness necessary for identifying security events.

\noindent\textbf{Event Clustering.}
\label{subsec:event_clustering_overview} 
With posts represented as dense embeddings, this stage groups semantically related discussions. As detailed in Section~\ref{subsec:event_clustering}, we employ an unsupervised clustering algorithm, specifically HDBSCAN~\cite{campello2013density}, to identify distinct event clusters \( \mathcal{C} \) by grouping proximal post embeddings in the vector space while designating noise points.

\noindent\textbf{Event Prioritization Ranking.}
Finally, to guide the analyst's attention to the most critical findings, the framework performs a ranking of all active event clusters. This ranking is based on a calculated Priority Score for each cluster, derived from metrics assessing its timeliness, relevance, credibility, and completeness~\cite{geras2024quality, chen2024stix}.

\subsection{Data Preparation}
\noindent\textbf{Post Classification.}
\begin{table*}[t!]
\renewcommand{\arraystretch}{1.1}
\caption{Distribution of categories in the forum post classification dataset with annotation guidelines.}
\label{tab:forum_categories} 
\footnotesize
\resizebox{0.95\textwidth}{!}{
\begin{tabular}{ll rll}
\toprule
\multicolumn{2}{l}{\textbf{Category}} & \textbf{\# Posts (Ratio)} & \textbf{Classification Criteria} & \textbf{Example Post Excerpt} \\
\midrule

\multicolumn{2}{l}{Irrelevant} & 5,925 (41.19\%) & \parbox[t]{5cm}{Forum posts unrelated to actionable security intelligence, including general discussions, advertisements, etc.} & \parbox[t]{6cm}{\textit{"Looking for pentesting team members with experience in wireless networks and physical security assessments."}} \\

\multicolumn{2}{l}{Security-Relevant} & 8,461 (58.81\%) & & \\

& DataBreach & 3,219 (22.38\%) & \parbox[t]{5cm}{Posts discussing unauthorized access to systems, exfiltration of sensitive information, or exposure of personal/corporate data.} & \parbox[t]{6cm}{\textit{"The database contains over 8.4M records including full names, email addresses, and hashed credentials from the compromised financial institution."}} \\

& Malware/Ransomware & 2,674 (18.59\%) & \parbox[t]{5cm}{Posts about malicious software, including detection, analysis, development, distribution, or ransomware activity.} & \parbox[t]{6cm}{\textit{"New loader bypasses Windows Defender by leveraging legitimate DLL sideloading technique and encrypting its C2 communications."}} \\

& Vulnerability & 722 (5.02\%) & \parbox[t]{5cm}{Posts revealing or discussing security weaknesses in software, hardware, or protocols, including zero-days and recently patched flaws.} & \parbox[t]{6cm}{\textit{"The buffer overflow in the authentication module can be triggered with a specially crafted request to achieve remote code execution."}} \\

& Fraud/Phishing & 816 (5.67\%) & \parbox[t]{5cm}{Posts related to social engineering attacks, credential harvesting campaigns, or financial scams.} & \parbox[t]{6cm}{\textit{"Selling fresh phishing kit that mimics the new Microsoft 365 login page with anti-bot protection and integrated Telegram notifications."}} \\

& DoS/DDoS & 890 (6.19\%) & \parbox[t]{5cm}{Posts discussing denial-of-service attacks, botnets, amplification techniques, or mitigation methods.} & \parbox[t]{6cm}{\textit{"Our booter service now supports layer 7 attacks with custom headers and bypass for common WAF implementations."}} \\

& BrandMonitoring & 140 (0.97\%) & \parbox[t]{5cm}{Posts about targeted threats, impersonation, leaked assets, or attacks linked to a specific brand/organization.} & \parbox[t]{6cm}{\textit{"Selling access to internal network of [Company Name], including employee credentials and source code."}} \\

\cline{1-3}
\multicolumn{2}{l}{\textbf{Total}} & \textbf{14,386 (100.0\%)} & & \\
\bottomrule
\end{tabular}
}
\end{table*}

CTI extraction increasingly relies on unstructured data from sources such as hacker forums. However, these platforms contain substantial non-relevant content, with actionable intelligence constituting only a small fraction of posts. This necessitates effective classification methodologies that extend beyond traditional keyword-based filtering~\cite{cui2024tweezers, paladini2024hackerforums}. Recent advances leverage Transformer-based models (e.g. BERT) to identify security-relevant content. While much prior work focused on binary relevance filtering~\cite{paladini2024hackerforums, fang2019analyzing, moreno2023cream}, this treats all relevant content as homogeneous. Recognizing the need for finer granularity, other approaches, similar to ours, have employed multi-class classification to assign posts to specific security categories (e.g., vulnerabilities, malware, data breaches) from a predefined set~\cite{deliu2018collecting, cui2024tweezers}. This level of detail, distinguishing between different types of security discussions, is critical for the accurate event clustering and meaningful feature construction used in downstream event prioritization, which are central goals of our work.

To support fine-grained threat differentiation, we perform multi-class classification by assigning each relevant post to its primary security category. To address the computational constraints of processing the full corpus, we first apply keyword-based filtering using terms drawn from prior research~\cite{shin2020cybersecurity, paladini2024hackerforums} and domain-specific vocabulary identified during preliminary exploration. From this filtered subset, we constructed a labeled dataset of 14,386 posts sampled from various forums within the CrimeBB corpus~\cite{Pastrana2018CrimeBB}. This dataset was manually annotated, with each post assigned to the most appropriate category based on its main topic. The annotation schema includes six categories informed by established cybersecurity taxonomies and common underground forum themes~\cite{cui2024tweezers, shin2020cybersecurity, sceller2017sonar, deliu2018collecting}: \textit{Irrelevant}, \textit{DataBreach}, \textit{Malware/Ransomware}, \textit{Vulnerability}, \textit{Fraud/Phishing}, \textit{DoS/DDoS}, and \textit{BrandMonitoring}. Table~\ref{tab:forum_categories} provides the distribution of these categories in our dataset, along with annotation guidelines and illustrative examples. The \textit{BrandMonitoring} category was specifically introduced to capture threats targeting specific organizations, such as attack plans, which are highly relevant to CTI but frequently underrepresented in technical taxonomies.

By assigning specific threat types, this categorization allows for more targeted analysis and noise reduction compared to basic relevance filtering.

\noindent\textbf{Named Entity Recognition (NER).} 
\label{subsec:ner_harmonization}
Following post classification, we apply Named Entity Recognition (NER) to extract structured information from forum posts. In the cybersecurity domain, NER identifies entities such as malware names, vulnerability identifiers, threat actors, tools, IP addresses, hashes, and targeted organizations~\cite{arazzi2023nlp, rahman2023what}. This structured extraction supports the core components of the \model{} framework, including both event clustering and prioritization.

Applying standard NER models to hacker forums is non-trivial due to domain-specific challenges. Although specialized cybersecurity NER datasets exist (for example, APTNER~\cite{wang2022aptner}, DNRTI~\cite{wang2020dnrti}), they are primarily derived from formal texts such as CTI reports and blogs. As a result, these models face two main limitations when transferred to forums.

\begin{enumerate}
   \item \textbf{Linguistic Disparity:} Forum language includes slang, misspellings, abbreviations, and informal syntax~\cite{paladini2024hackerforums}. Models trained on formal language often fail to recognize entities expressed in this irregular form.

   \item \textbf{Information Granularity:} Forum content frequently contains operational details, such as payloads, compromised databases, or specific attack vectors. Existing NER schemas typically abstract these details, focusing instead on high-level threat indicators. Figure \ref{fig:ner_example} provides an example that highlights the extraction of such operational entities from a data breach announcement.
\end{enumerate}

To address these limitations and avoid the cost of developing large annotated forum-specific corpora, we explore zero-shot NER using open Large Language Models (LLMs). Previous work shows that LLMs, when guided by prompt engineering, can yield competitive results on NER tasks~\cite{cheng2024novel, ashok2023promptner}, including in few-shot and zero-shot settings. Their pretraining on diverse corpora enables them to generalize across writing styles, making them well-suited for processing informal forum content.

The entities extracted via NER serve two primary functions within our pipeline. First, they provide essential signals for training the entity-aware dense embedding model, enabling it to capture event-specific semantics more effectively and link posts containing shared indicators. Second, the extracted entities and their associated metadata (e.g., entity type, frequency) are used by the prioritization module to assess event characteristics such as topical relevance and potential impact. 

\subsection{Post Representation}
\label{subsec:post_representation}

A critical step for effective event clustering is transforming raw hacker forum posts into meaningful vector representations that capture their core content and relatedness in the security domain. Our objective is to map each post \( p \) to a single, dense vector \( \mathbf{h}_p \in \mathbb{R}^d \) such that posts discussing the same underlying security event are mapped to vectors that are proximal in the embedding space, while vectors for posts about different events are distant.

\noindent\textbf{Entity-Aware Dense Embeddings.}
Dense contextual embeddings, generated by large pre-trained Transformer models like BERT~\cite{devlin2019bert}, excel at capturing rich semantic meaning and contextual nuances. Leveraging this capability, these embeddings form the foundation of our post representation strategy. However, standard Transformer embeddings distribute attention relatively uniformly. This can underemphasize the crucial signal carried by specific, often rare, domain-specific Named Entities (NEs)—like unique CVE IDs, malware hashes, or threat actor handles—which are fundamental identifiers for distinguishing concrete security events~\cite{saravanakumar2021event, staykovski2019dense}. Consequently, relying solely on standard dense representations may make it difficult to effectively differentiate posts discussing related concepts generally from those pertaining to a specific, actionable event defined by key NEs.

\begin{figure}[t]
    \centering
    
    \includegraphics[width=\columnwidth]{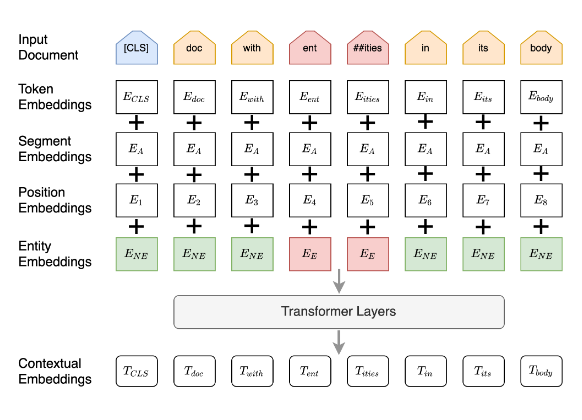}
    \caption{Illustration of the entity-aware BERT model architecture, showing the addition of entity presence embeddings to the standard input embeddings before the Transformer layers.}
    \label{fig:dense_embeddings_archi} 
    \vspace{-5pt}
\end{figure}

To explicitly enhance the representation's sensitivity to these critical identifiers, we employ an entity-aware BERT model (illustrated in Figure~\ref{fig:dense_embeddings_archi}).Inspired by approaches integrating entity information into representations for related tasks~\cite{saravanakumar2021event, logeswaran2019zero}, we adapt the standard Transformer architecture to integrate information about identified NEs. As extracted in Section~\ref{subsec:ner_harmonization}, each token in a post is labeled based on whether it belongs to a recognized security-relevant entity. This entity presence information is then incorporated directly into the model's input layer. We train a dedicated entity presence embedding layer, mapping tokens to distinct embeddings based on their entity status (e.g., belonging to an entity, not belonging to an entity, or padding). During the forward pass, the standard input embeddings (token, segment, and position embeddings from the base Transformer) are augmented by summing them with the corresponding entity presence embedding for each token, \emph{before} feeding the composite embedding into the Transformer encoder layers:
\[ \mathbf{E}_{\text{input}} = \mathbf{E}_{\text{token}} + \mathbf{E}_{\text{segment}} + \mathbf{E}_{\text{position}} + \mathbf{E}_{\text{entity presence}} \]
This input-level augmentation encourages the model's internal representation process to give specific consideration to tokens marked as part of an entity, without requiring changes to the core Transformer layers themselves. After the Transformer encoder, the final hidden states are pooled using the CLS token representation, and then passed through an MLP layer to produce the fixed-size dense post embedding vector \( \mathbf{h}_p \).

\begin{figure}[t]
    \centering

    \includegraphics[width=0.8\columnwidth]{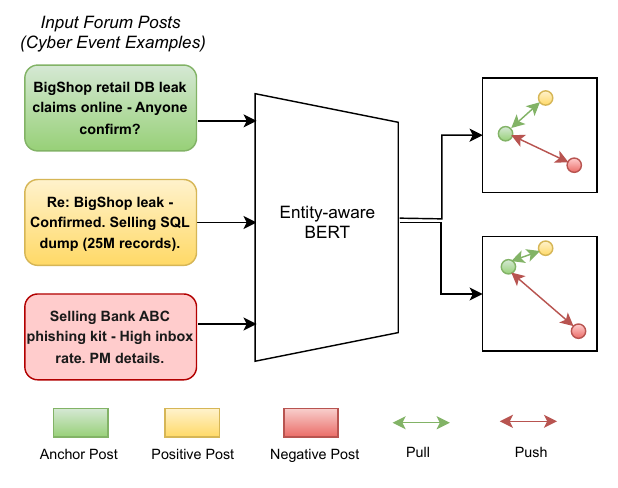} 
    \caption{Illustration of the combined contrastive fine-tuning process for the entity-aware BERT model. The model is trained using Post Triplet (\(\mathcal{L}_t\)) and Post Pairwise (\(\mathcal{L}_p\)) losses to structure the embedding space such that embeddings of posts from the same event are closer than those from different events.}
    \label{fig:dense_embeddings_training} 
    \vspace{-5pt}
\end{figure}

\noindent\textbf{Contrastive Fine-Tuning Objective.}
To further ensure the resulting entity-aware embeddings are optimally structured for event clustering—where proximity in the vector space directly corresponds to event relatedness—we fine-tune the model using a combined contrastive learning objective. Contrastive learning aims to pull representations of "positive" pairs (e.g., posts from the same event) closer while pushing "negative" pairs (e.g., posts from different events) apart. Our approach integrates two complementary post-level contrastive loss functions. Figure~\ref{fig:dense_embeddings_training} illustrates this fine-tuning process. 

The first component is a Post Triplet Margin Loss (\(\mathcal{L}_t\)). This loss operates on triplets of (anchor post, positive post, negative post), where the positive post belongs to the same ground-truth event as the anchor, and the negative post belongs to a different ground-truth event. The objective is to ensure that the embedding of the anchor post (\(\mathbf{h}_a\)) is closer in the embedding space (using cosine similarity \( \text{sim}(\cdot, \cdot) \), where higher similarity implies smaller effective distance) to the embedding of the positive post (\(\mathbf{h}_p\)) than it is to the embedding of the negative post (\(\mathbf{h}_n\)), by at least a defined margin \( \alpha \). The loss is formulated as:
\begin{equation}
\label{eq:triplet_loss_rev}
\mathcal{L}_t = \max \left( \text{sim}(\mathbf{h}_a, \mathbf{h}_n) - \text{sim}(\mathbf{h}_a, \mathbf{h}_p) + \alpha, 0 \right)
\end{equation}
Minimizing this loss pulls \( \mathbf{h}_a \) and \( \mathbf{h}_p \) together while pushing \( \mathbf{h}_a \) and \( \mathbf{h}_n \) apart based on a sampled hard negative post.

The second component is a Post Pairwise Cosine Loss (\(\mathcal{L}_p\)), inspired by SimCSE~\cite{gao2021simcse}. This loss leverages in-batch negatives. For a batch of \( N \) anchor-positive pairs (derived from \( N \) distinct original posts), resulting in \( 2N \) post embeddings (\(\mathbf{h}_{a,1}, \dots, \mathbf{h}_{a,N}\) and \(\mathbf{h}_{p,1}, \dots, \mathbf{h}_{p,N}\)), the similarity matrix \( \mathbf{S} \in \mathbb{R}^{N \times N} \) is computed where \( S_{ij} = \text{sim}(\mathbf{h}_{a,i}, \mathbf{h}_{p,j}) / \tau \), scaled by a temperature parameter \( \tau \). The loss is then the standard cross-entropy between this similarity matrix and the identity matrix, effectively training the model to predict that \( \mathbf{h}_{a,i} \) corresponds to its positive pair \( \mathbf{h}_{p,i} \) among all positive embeddings in the batch:
\begin{equation}
\label{eq:pairwise_loss_rev}
\mathcal{L}_p = - \frac{1}{N} \sum_{i=1}^N \log \frac{e^{\text{sim}(\mathbf{h}_{a,i}, \mathbf{h}_{p,i}) / \tau}}{\sum_{j=1}^N e^{\text{sim}(\mathbf{h}_{a,i}, \mathbf{h}_{p,j}) / \tau}}
\end{equation}
This loss encourages intra-pair compactness and inter-pair separation using a dynamic set of negatives within each batch.

The overall fine-tuning objective is a weighted sum of these two post-level contrastive loss components: \( \mathcal{L}_{\text{total}} = w_t \mathcal{L}_t + w_p \mathcal{L}_p \), where \( w_t, w_p \) are configurable weights. This combined strategy leverages the complementary strengths of hard negative sampling (\(\mathcal{L}_t\)) and in-batch negative sampling (\(\mathcal{L}_p\)) to produce dense, entity-aware post embeddings that effectively capture event relatedness for the subsequent clustering phase.

\subsection{Event Clustering}
\label{subsec:event_clustering}

With each post transformed into a dense, entity-aware embedding \( \mathbf{h}_p \), the next step is to group these representations into clusters that correspond to discrete security events discussed across the forum corpus (\( f_{\text{cluster}} \) in the Problem Statement). Given the unsupervised nature of the problem—we do not know the number or exact boundaries of events beforehand—and the potential for noise and variations in cluster density within real-world forum data, we require a robust clustering algorithm capable of discovering arbitrary cluster shapes and effectively identifying noise points.

Based on these requirements, we employ Hierarchical Density-Based Spatial Clustering of Applications with Noise (HDBSCAN)~\cite{campello2013density}. HDBSCAN is a powerful density-based clustering algorithm. Unlike partition-based methods, HDBSCAN transforms the space into a hierarchy of density-based clusters. It then uses a technique to extract a flat partitioning from this hierarchy. This unique capability allows HDBSCAN to discover clusters of varying densities within the data and is particularly effective at distinguishing core clusters from background noise.

In our framework, after generating the dense embeddings for a set of posts, we apply the HDBSCAN algorithm to the embedding space. The clustering algorithm is configured to group a minimum number of related posts considered sufficient to constitute a potential "event" cluster. Posts that the algorithm does not assign to any cluster are treated as noise.

The output of this clustering process is the set of discovered event clusters \( \mathcal{C} = \{c_1, c_2, \ldots, c_m\} \). Each cluster \( c_i \) comprises the post IDs whose embeddings were grouped together by the algorithm. These clusters then serve as the input for the subsequent event prioritization phase, enabling us to rank the discovered events based on their estimated operational significance.

\subsection{Event Prioritization Ranking}
\label{subsec:event_ranking}

A core contribution of \model{} is its ability not only to detect security event clusters but also to prioritize them, guiding security analysts towards the most critical or impactful ongoing events discussed within hacker forums. Drawing from established CTI quality frameworks~\cite{tundis2022feature, geras2024quality, zibak2022threat}, our prioritization mechanism is designed to integrate four key dimensions operationally relevant to security analysts: Timeliness (T), Relevance (V), Credibility (R), and Completeness (C). This score, calculated daily as a \textbf{Priority Score} \( s_i \) for each active event cluster \( c_i \in \mathcal{C} \), is designed to reflect the potential operational importance of the event.

We quantify each dimension for a given cluster \( c_i \) based on properties derived from its constituent posts \( \{p \mid p \in c_i\} \), associated metadata from the CrimeBB schema (e.g., member reputation, post timestamps), and entities extracted by our NER component.

\noindent\textbf{Timeliness (T):} This metric captures both the recency and the level of activity associated with the event cluster. It combines two components:
\begin{itemize}
    \item \textit{Recency (\(Rec\))}: Measures how recently the latest post in the cluster appeared, using an exponential decay function:
    \begin{equation}
        Rec(c_i) = \exp\left(-\frac{t_{\text{now}} - t_{c_i}^{\max}}{\tau_{Rec}}\right)
        \label{eq:recency}
    \end{equation}
    where \( t_{\text{now}} \) is the current time, \( t_{c_i}^{\max} \) is the timestamp of the latest post in cluster \( c_i \), and \( \tau_{Rec} \) is a hyperparameter controlling the decay rate.
    \item \textit{Activity (\(Act\))}: Reflects the volume of discussion within the cluster, measured by the logarithm of the number of posts \( N(c_i) \) in the cluster:
    \begin{equation}
        Act(c_i) = \log(1 + N(c_i))
        \label{eq:activity}
    \end{equation}
\end{itemize}
The final Timeliness score is a weighted combination:
\begin{equation}
    T(c_i) = w_{Rec} \cdot Rec(c_i) + w_{Act} \cdot Act(c_i)
    \label{eq:timeliness}
\end{equation}
where \( w_{Rec} \) and \( w_{Act} \) are non-negative weights. This formulation prioritizes events that are both recent and have generated significant discussion.

\noindent\textbf{Relevance (V):} This metric assesses the cluster's relevance to specific analyst interests, which can be represented by an optional query \( Q \). When a query is provided, the relevance \( V(c_i, Q) \) is calculated based on the semantic similarity between the cybersecurity entities present in the cluster and those specified or implied by the query. It is computed as:
\begin{equation}
    V(c_i, Q) = f_{\text{match}}(\text{Entities}(c_i), \text{Entities}(Q))
    \label{eq:relevance}
\end{equation}
where \( \text{Entities}(c_i) \) denotes the set of unique cybersecurity entities extracted from the posts in cluster \( c_i \) (via NER, Section~\ref{subsec:ner_harmonization}), and \( \text{Entities}(Q) \) represents the set of target entities derived from the analyst query \( Q \). The function \( f_{\text{match}} \) specifically calculates the semantic similarity between these two sets of entities. If no query \( Q \) is provided by the analyst, this component defaults to zero, i.e., \( V(c_i, \emptyset) = 0 \).

\noindent\textbf{Credibility (R):} This metric estimates the reliability of the information within the cluster based on the reputation or standing of the contributing authors within the forum community. Let \( \mathcal{A}(c_i) \) denote the set of unique authors who contributed posts to cluster \( c_i \). The credibility score is computed as the average reputation score across these unique authors:
\begin{equation}
    R(c_i) = \frac{1}{|\mathcal{A}(c_i)|} \sum_{a \in \mathcal{A}(c_i)} \text{Reputation}(a)
    \label{eq:credibility}
\end{equation}
where \( |\mathcal{A}(c_i)| \) is the number of unique authors in the cluster. The \texttt{Reputation(a)} for each author is derived directly from user metadata available in the CrimeBB schema, which include metrics such as post count, join date, or explicit reputation points awarded by other forum members. A higher average reputation score among the cluster's contributors suggests potentially more credible or influential discussions.

\noindent\textbf{Completeness (C):} This metric evaluates the richness and diversity of information contained within the event cluster, based on the variety and quantity of extracted cybersecurity entities. It is calculated as:
\begin{equation}
\label{eq:completeness}
\begin{split}
C(c_i) = &\ w_\alpha \cdot \log(1 + |\text{unique\_entities}(c_i)|) \\
         & + w_\beta \cdot \log(1 + |\text{unique\_entity\_types}(c_i)|)
\end{split}
\end{equation}
where \( |\text{unique\_entities}(c_i)| \) is the count of unique entity instances (e.g., specific CVEs, malware hashes) extracted from posts in \( c_i \) by the NER component, \( |\text{unique\_entity\_types}(c_i)| \) is the count of distinct entity types (e.g., 'Vulnerability', 'Malware', 'Threat-Actor') present among the extracted entities, and \( w_{\alpha}, w_{\beta} \) are weighting parameters. A higher score indicates a more developed event narrative with diverse indicators.

\subsection{Priority Score Calculation}
\label{subsec:priority_score_calc} 

The final Priority Score \( s_i \) for an event cluster \( c_i \) is determined by aggregating the normalized scores [0, 1] from the four key dimensions: Timeliness (\(T\)), Relevance (\(V\)), Credibility (\(R\)), and Completeness (\(C\)). This aggregation is performed using a weighted linear combination:
\begin{equation}
    s_i(Q) = w_{T} T(c_i) + w_{V} V(c_i, Q) + w_{R} R(c_i) + w_{C} C(c_i)
    \label{eq:priority_score_final} 
\end{equation}
Here, the coefficients \( w_{T}, w_{V}, w_{R}, w_{C} \) represent non-negative weights assigned based on the perceived importance of each dimension for operational CTI. These weights are configurable and can be optimized based on empirical analysis or direct analyst input to reflect specific monitoring priorities.

Finally, on a regular basis, the active event clusters \( c_i \in \mathcal{C} \) are ranked in descending order based on their calculated Priority Scores \( s_i(Q) \), generating the prioritized list \( \mathcal{R} \) intended for analyst review. This allows analysts to focus on the clusters deemed most operationally significant according to the defined criteria.

\begin{table}[t!]
\caption{Characteristics of Analyzed Hacker Forums}
\label{tab:forum_stats}
\centering
\resizebox{\columnwidth}{!}{%
\begin{tabular}{lrrrrl}
\toprule
\textbf{Forum} & \textbf{Posts} & \textbf{Threads} & \textbf{Members} & \textbf{Boards} & \textbf{Time Span} \\
\midrule
HackForums & 42,474,325 & 4,148,196 & 716,058 & 212 & Jan 2007 - Apr 2023 \\
Nulled & 6,675,497 & 591,830 & 1,647,057 & 168 & Apr 2013 - Jun 2023 \\
Cracked & 2,977,800 & 419,517 & 897,760 & 163 & Apr 2018 - Jun 2023 \\
BreachForums & 737,922 & 34,412 & 119,260 & 72 & Mar 2022 - Mar 2023 \\
KernelMode & 26,815 & 3,606 & 1,668 & 11 & Mar 2010 - Nov 2019 \\
\bottomrule
\end{tabular}
}
\end{table}

\section{Experiments and Evaluation} 
\label{sec:experiments}
To rigorously evaluate the performance of the \model{} framework in identifying and prioritizing security events within complex hacker forum data, we conducted a comprehensive set of experiments. This section details the experimental methodology, including the datasets used, the specific configurations tested, and the metrics employed to assess each stage of the framework, from data preparation to final event ranking.

\subsection{Experimental Setup}
\label{subsec:exp_setup}

\noindent\textbf{Dataset Description:}
\begin{table*}[t!]
\footnotesize
\renewcommand{\arraystretch}{1.1}
\caption{
Performance comparison of transformer models on forum post classification. Results reported as F1 scores for each category.
}
\label{tab:forum_classification_results}
\begin{adjustbox}{width=\textwidth, center}
\begin{tabular}{ccccccccc}
\toprule
& Irrelevant & DataBreach & Malware & Vulnerability & FraudPhishing & DosAttack & BrandMonitoring & Avg. F1 \\
\midrule
BERT           & 0.7266 & \underline{0.9367} & 0.8762 & 0.7326 & \underline{0.5449} & \textbf{0.9003} & 0.2632 & 0.7115 \\
RoBERTa        & \underline{0.7478} & \textbf{0.9478} & 0.8721 & 0.7397 & 0.5359 & 0.8920 & 0.2727 & 0.7154 \\
SecureBERT     & 0.7548 & 0.9414 & 0.8698 & \textbf{0.7755} & 0.4828 & 0.8857 & \textbf{0.4091} & \underline{0.7313} \\
CySecBERT      & \textbf{0.7559} & 0.9455 & \textbf{0.8844} & 0.7529 & 0.5231 & \textbf{0.9003} & 0.3043 & 0.7238 \\
DarkBERT       & 0.7540 & 0.9399 & \underline{0.8841} & \underline{0.7626} & \textbf{0.5460} & 0.8952 & \underline{0.3902} & \textbf{0.7389} \\
\bottomrule
\end{tabular}
\end{adjustbox}
\end{table*} 

Our empirical evaluation leverages data from five prominent English-language forums within the CrimeBB dataset~\cite{Pastrana2018CrimeBB}: \textit{Nulled} and \textit{Cracked}\footnote{Major cybercrime forums taken down in Jan 2025 post-collection.}, \textit{BreachForums}\footnote{Known for trading breached data/exploits; data precedes takedowns.}, \textit{KernelMode}\footnote{Technical discussions on malware, system internals, vulnerabilities.}, and \textit{HackForums}\footnote{Long-running platform covering diverse hacking topics.}.

These platforms were specifically selected due to their central roles within the cybercriminal ecosystem and their rich, diverse discussions spanning malware development, vulnerability exploitation, data breaches, and advanced hacking methodologies. As detailed in Table~\ref{tab:forum_stats}, our dataset encompasses millions of posts across multiple years, providing a robust and representative cross-section of contemporary underground security discourse. This carefully curated data selection enables rigorous assessment of EventHunter's performance in accurately identifying and prioritizing significant security events from extensive and noisy forum interactions.

\noindent\textbf{Evaluation Data \& Ground Truth:}
Evaluating unsupervised clustering algorithms, particularly in specialized domains like cybersecurity event detection from noisy forum data, necessitates the creation of reliable ground truth. Standard labeled datasets for fine-grained security \textit{events} spanning multiple fragmented posts are largely unavailable. Therefore, to assess the ability of \model{} to group related posts into coherent event clusters, we constructed a ground truth dataset based on well-documented, distinct security incidents discussed within the forums used in our study.

Specifically, we identified a set of 70 known, significant security events that generated traceable discussions within our forum corpus. These included incidents such as major vulnerability disclosures (e.g., CVE-2022-42475, a critical FortiOS RCE vulnerability), specific ransomware campaigns (e.g., discussions surrounding LockBit 3.0 operations), and widely publicized data breaches (e.g., the leak involving the Turkish Public Health System, HSYS). For each such reference event, we meticulously curated a ground truth cluster by: (1) Identifying posts definitively discussing that specific event using unique and unambiguous markers (e.g., canonical vulnerability identifiers like `CVE-2022-42475', established malware family names, specific targeted entity names like `HSYS' coupled with incident context) within a relevant time window. (2) Verifying through \textbf{manual inspection} that the collected posts genuinely pertain to the same underlying security incident and were not merely tangential mentions.

This process yielded a collection of disjoint sets of forum posts, where each set represents a single, distinct real-world security event (e.g., one set for all posts discussing CVE-2022-42475, another for the HSYS leak). This curated collection serves as the ground truth against which we evaluate the clustering component of \model{}. While acknowledging the inherent challenges in creating exhaustive ground truth for dynamic events, this approach provides a robust benchmark based on verifiable real-world occurrences discussed in the forums.

\noindent\textbf{Implementation Details:}
We implemented the \model{} framework using Python (version 3.8), leveraging core libraries including PyTorch~\cite{paszke2019pytorch} for deep learning models and scikit-learn~\cite{pedregosa2011scikit} for traditional methods and evaluation metrics. The embedding size for all Transformer models was 768. For clustering, HDBSCAN was configured with a \texttt{min\_cluster\_size} of 5. To ensure reproducibility, key components were configured as follows: Transformer models (used for post classification and generating dense embeddings, e.g., DarkBERT, RoBERTa) were fine-tuned using the AdamW optimizer~\cite{loshchilov2019decoupled} with a learning rate of \(2 \times 10^{-5}\), a dropout rate of 0.1, and trained for 5 epochs. The margin \(\alpha\) for the contrastive triplet loss component was set to 0.5, a common value in the contrastive learning literature. TF-IDF vectors were generated using standard scikit-learn configurations. The LLM-based NER component utilized zero-shot prompting via the Ollama library\footnote{\url{https://ollama.com/}} with the \textit{mistral-nemo} model. The source code for this paper is available at: \url{https://github.com/yasirech-chammakhy/EventHunter}.

\noindent\textbf{Evaluation Metrics:}
We evaluate clustering performance using five standard metrics: Adjusted Rand Index (ARI)~\cite{hubert1985comparing}, which measures the similarity between two clusterings while adjusting for chance; Normalized Mutual Information (NMI)~\cite{vinh2010information}, which quantifies the mutual dependence between the predicted and true clusters. These metrics range from 0 to 1, with higher values indicating better clustering quality. Our implementation uses the scikit-learn library's clustering metrics module for consistent and reproducible evaluation.

\subsection{Multi-Category Classification Model Evaluation}
\label{subsec:classification_eval}

To focus downstream analysis on relevant content, \model{} first employs a multi-category classification step. We evaluated several pre-trained Transformer models for this task: \textit{BERT-base}~\cite{devlin2019bert}, \textit{RoBERTa}~\cite{liu2019roberta}, \textit{SecureBERT}~\cite{aghaei2022securebert}, \textit{CySecBERT}~\cite{fiorini2023cysecbert}, and \textit{DarkBERT}~\cite{jin2023darkbert}. Each model was fine-tuned using a standard architecture comprising the base Transformer followed by a classification head (dropout and linear layer on the [CLS] token representation) predicting one of the seven security-relevant categories identified in Table~\ref{tab:forum_categories}. 

Training utilized a weighted cross-entropy loss function to mitigate class imbalance, with optimizer, learning rate, and other hyperparameters detailed in Section~\ref{subsec:exp_setup}. Table~\ref{tab:forum_classification_results} summarizes the classification performance. Consistent with expectations for domain-specific text, models pre-trained on cybersecurity corpora (DarkBERT, CySecBERT) demonstrated superior effectiveness. DarkBERT achieved the highest average F1 score across categories and was therefore selected as the classification component for the \model{} pipeline. This step yields a filtered and categorized set of posts pertinent to security discussions, preparing the data for representation learning and event clustering.

\subsection{NER Performance}
\label{subsec:ner_performance}

Named Entity Recognition (NER) is crucial for \model{} to identify key actors, tools, vulnerabilities, and targets that anchor security events within forum discussions. Extracted entities inform both representation learning and event prioritization. To select an effective NER component without requiring extensive fine-tuning, we evaluated several recent open Large Language Models (LLMs) using zero-shot prompting on the standard CyNER dataset~\cite{alam2022cyner}. Specifically, we assessed \textit{gemma2}\footnote{\url{https://blog.google/technology/developers/google-gemma-2/}}, \textit{mistral-nemo}\footnote{\url{https://mistral.ai/news/mistral-nemo}}, and \textit{mistral:7b-instruct}\footnote{\url{https://mistral.ai/news/announcing-mistral-7b}}.

Performance was measured by mention-level F1 score. \textit{Mistral-nemo} achieved the highest F1 (52.13), outperforming \textit{gemma2} (49.39) and \textit{mistral:7b-instruct} (42.10). None of these LLMs were fine-tuned on CyNER or any cybersecurity-specific corpus prior to this zero-shot evaluation. Notably, this zero-shot performance surpasses the supervised RoBERTa-base benchmark reported in~\cite{cui2024tweezers} (F1 score of 39.7), despite using no task-specific supervision. Consequently, we adopt \textit{mistral-nemo} with zero-shot prompting, aligned to our predefined entity schema, as the default NER component in \model{}. The entities extracted by this component serve key roles within the pipeline, providing crucial signals for representation learning and metadata for event prioritization.

\footnotetext[1]{\url{https://blog.google/technology/developers/google-gemma-2/}}
\footnotetext[2]{\url{https://mistral.ai/news/mistral-nemo}}
\footnotetext[3]{\url{https://mistral.ai/news/announcing-mistral-7b}}

\begin{table}[tbp] 
\centering
\caption{Finding 1: Transformer Embeddings vs. Baselines (Best NMI per Transformer model type shown, using standard contrastive loss)}
\label{tab:finding1_transformers_vs_baselines}
\resizebox{\columnwidth}{!}{
\begin{tabular}{lccc}
\toprule
\textbf{Model / Method}           & \textbf{ARI}    & \textbf{NMI}     & \textbf{Clusters Found} \\
\midrule
\textit{Baselines} \\
TF-IDF                   & 0.264  & 0.638   & 12 \\
Word2Vec (GloVe)         & -0.024 & 0.222   & 4 \\
\midrule
\textit{Transformers (Best NMI - Standard Contrastive Loss)} \\
BERT (Pairwise)     & 0.159  & 0.534   & 9 \\
DarkBERT (Pairwise) & 0.347  & 0.712   & 14 \\
RoBERTa (Triplet+Pairwise)  & 0.402  & 0.714   & 11 \\
CySecBERT (Pairwise)& 0.377 & 0.689  & 12 \\
\bottomrule
\end{tabular}
} 
\vspace{-5pt} 
\end{table}

\subsection{Clustering Results}
\label{subsec:clustering_results}

We evaluated the effectiveness of \model{}'s event clustering component using the ground truth dataset derived from the curation process described in Section~\ref{subsec:exp_setup}. From the initially identified set of 70 distinct security events, we randomly selected a subset of 21 events to serve as the test set for this clustering evaluation. This curated test set, representing a diverse range of incidents discussed in the forums, served as the ground truth against which we assessed how well different post embedding configurations group related posts. While this test set provides a valuable benchmark for relative comparison, evaluating performance against larger-scale event datasets remains an area for future work.

\textbf{Finding 1: Fine-tuned Transformer embeddings significantly outperform traditional baselines.}
Learned contextual embeddings offer significant gains over traditional vectorization approaches. As shown in Table~\ref{tab:finding1_transformers_vs_baselines}, TF-IDF and Word2Vec underperform (NMI 0.638 and 0.222 respectively), while RoBERTa and DarkBERT achieve substantially higher scores (NMI 0.714 and 0.712), validating the impact of deep contextual representation and contrastive fine-tuning.

\textbf{Finding 2: Domain-Specific Pre-training Enhances Clustering.}
Comparing the best configurations for each Transformer architecture (Table~\ref{tab:finding2_domain_vs_general}, using Pairwise Loss results for comparison) highlights the benefit of domain adaptation. Models pre-trained on relevant corpora (DarkBERT, CySecBERT) consistently outperform general-purpose models (BERT, RoBERTa). DarkBERT achieves the highest NMI, while CySecBERT attains the highest ARI. This indicates that familiarity with cybersecurity jargon, topics, and linguistic patterns acquired during pre-training enables these models to learn more discriminative embeddings for forum data.

\begin{table}[tbp] 
\centering
\caption{Finding 2: Domain-Specific vs. General Models (Best Standard Contrastive Loss configuration per model from Input entity experiments)}
\label{tab:finding2_domain_vs_general}
\resizebox{\columnwidth}{!}{
\begin{tabular}{lccc}
\toprule
\textbf{Model Type (Best Standard Loss)} & \textbf{ARI}    & \textbf{NMI}    & \textbf{Clusters Found} \\
\midrule
\textit{General Domain} \\
BERT (Pairwise)         & 0.159  & 0.534  & 9 \\
RoBERTa (Pairwise)      & 0.314  & 0.664  & 13 \\
\midrule
\textit{Cybersecurity Domain} \\
DarkBERT (Pairwise)     & 0.347  & \textbf{0.712} & 14 \\
CySecBERT (Pairwise)    & \textbf{0.377}  & 0.689  & 12 \\
\bottomrule
\end{tabular}
} 
\vspace{-5pt} 
\end{table}

\begin{figure}[tbp] 
\centering
\includegraphics[width=\columnwidth]{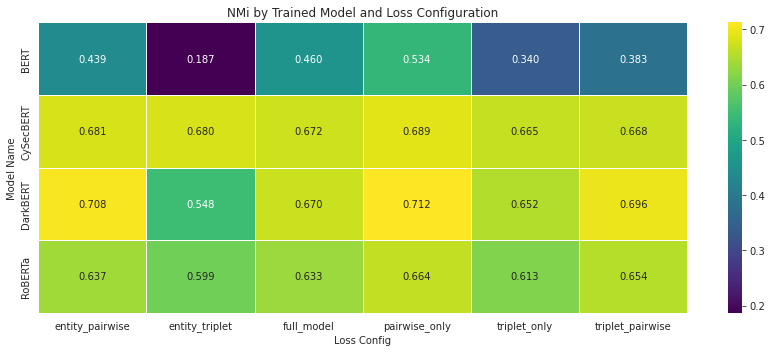}
\caption{NMI Scores by Model and Loss Configuration (Input Entity Experiments).}
\label{fig:nmi_heatmap}
\vspace{-10pt} 
\end{figure}

\textbf{Finding 3: Pairwise or combined contrastive objectives outperform triplet-only loss.}
Contrastive loss formulations significantly influence clustering performance. Table~\ref{tab:finding3_loss_comparison} and the NMI heatmap in Figure~\ref{fig:nmi_heatmap} show that triplet-only loss generally underperforms compared to pairwise or combined losses across models (using standard inputs, from Input entity experiments). For instance, DarkBERT achieves NMI 0.652 (triplet), 0.712 (pairwise), and 0.696 (combined), reinforcing the value of in-batch negative sampling over hard negative mining alone.

\begin{table}[tbp] 
\centering
\caption{Finding 3: Comparison of Standard Contrastive Losses (Input Entity Experiments)}
\label{tab:finding3_loss_comparison}
\resizebox{\columnwidth}{!}{

\begin{tabular}{lcccc}
\toprule
\textbf{Model}     & \textbf{Loss Configuration} & \textbf{ARI}    & \textbf{NMI}     & \textbf{Clusters Found} \\ 
\midrule
      & Triplet Loss       & 0.083  & 0.340   & 6 \\ 
BERT      & Pairwise Loss      & \textbf{0.159}  & \textbf{0.534}   & 9 \\ 
      & Combined Loss      & 0.112  & 0.383   & 5 \\ 
\midrule
  & Triplet Loss       & 0.240  & 0.652   & 12 \\ 
DarkBERT  & Pairwise Loss      & \textbf{0.347}  & \textbf{0.712}   & 14 \\ 
  & Combined Loss      & 0.303  & 0.696   & 12 \\ 
\midrule
   & Triplet Loss       & 0.319  & 0.613   & 8 \\ 
RoBERTa   & Pairwise Loss      & 0.314  & \textbf{0.664}   & 13 \\ 
   & Combined Loss      & \textbf{0.342}  & 0.654   & 9 \\ 
\midrule
 & Triplet Loss       & 0.328  & 0.665   & 10 \\ 
CySecBERT & Pairwise Loss      & \textbf{0.377}  & \textbf{0.689}   & 12 \\ 
 & Combined Loss      & 0.327  & 0.668   & 11 \\ 
\bottomrule
\end{tabular}
} 
\vspace{-5pt} 
\end{table}

\textbf{Finding 4: Input-Level Entity-Aware Mechanism Shows Model-Specific Benefits.}
We evaluated an input-level entity-aware mechanism, modifying input embeddings with entity indicators, against a standard contrastive baseline (Table~\ref{tab:finding4_entity_impact}). The results ('Input Entity+' vs. 'Standard Input') show benefits that depend on the base model and loss function. Notably, the mechanism significantly improved BERT performance with Combined Loss (NMI increased from 0.383 to 0.460) and yielded competitive results with DarkBERT (achieving higher ARI under Pairwise and Combined losses, e.g., 0.368 vs. 0.347 with Pairwise). However, the standard baseline generally performed better for RoBERTa and CySecBERT in our tested configurations (e.g., RoBERTa NMI 0.664 vs. 0.637 with Pairwise loss). While not universally superior in these tests, the input-level entity-aware approach demonstrates clear potential depending on the specific model and setup, suggesting a promising direction for further tuning.

\begin{table*}[tp]
\centering

\caption{Finding 4: Performance Comparison: Standard Input vs. Input-Level Entity-Aware Mechanism}
\label{tab:finding4_entity_impact}
\begin{tabular}{l l cc cc cc}
\toprule
& & \multicolumn{2}{c}{\textbf{Triplet Loss}} & \multicolumn{2}{c}{\textbf{Pairwise Loss}} & \multicolumn{2}{c}{\textbf{Combined Loss (Full)}} \\
\cmidrule(lr){3-4} \cmidrule(lr){5-6} \cmidrule(lr){7-8}
\textbf{Model} & \textbf{Mechanism} & \textbf{ARI} & \textbf{NMI} & \textbf{ARI} & \textbf{NMI} & \textbf{ARI} & \textbf{NMI} \\
\midrule
\multirow{2}{*}{BERT}
& Standard Input & 0.083 & 0.340 & 0.159 & 0.534 & 0.112 & 0.383 \\
& Input Entity+  & 0.050 & 0.187 & 0.059 & 0.439 & \textbf{0.230} & \textbf{0.460} \\
\midrule
\multirow{2}{*}{DarkBERT}
& Standard Input & 0.240 & 0.652 & 0.347 & \textbf{0.712} & 0.303 & 0.696 \\
& Input Entity+  & 0.206 & 0.548 & \textbf{0.368} & 0.708 & 0.31 & 0.670 \\ 
\midrule
\multirow{2}{*}{RoBERTa}
& Standard Input & 0.319 & 0.613 & 0.314 & \textbf{0.664} & \textbf{0.342} & 0.654 \\
& Input Entity+  & 0.278 & 0.599 & 0.217 & 0.637 & 0.286 & 0.633 \\
\midrule
\multirow{2}{*}{CySecBERT}
& Standard Input & 0.328 & 0.665 & \textbf{0.377} & \textbf{0.689} & 0.327 & 0.668 \\
& Input Entity+  & 0.335 & 0.680 & 0.279 & 0.681 & 0.330 & 0.672 \\
\bottomrule
\end{tabular}
\vspace{2pt}

\vspace{-5pt}
\end{table*}

\textbf{Finding 5: Output-Level Entity-Aware Loss Did Not Yield Benefits in Tested Configurations.}

\label{sec:finding5}
As an alternative entity-aware strategy, we explored augmenting the standard post-level contrastive objectives with an auxiliary \textit{output-level entity contrastive loss} (\(\mathcal{L}_e\)). This loss aims to make the final post embedding \(\mathbf{h}_p\) explicitly aware of the named entities within the post by pulling it closer to embeddings of mentioned entities (\(e^+\)) and pushing it away from embeddings of unmentioned entities (\(e^-\)). We evaluated adding this entity loss (\(\mathcal{L}_e\), weighted by \(w_e\)) to the standard objectives (\(\mathcal{L}_t + w_e \mathcal{L}_e\), \(\mathcal{L}_p + w_e \mathcal{L}_e\), \(\mathcal{L}_t + \mathcal{L}_p + w_e \mathcal{L}_e\)).

\begin{figure}[htbp]
    \centering
    \includegraphics[width=\columnwidth]{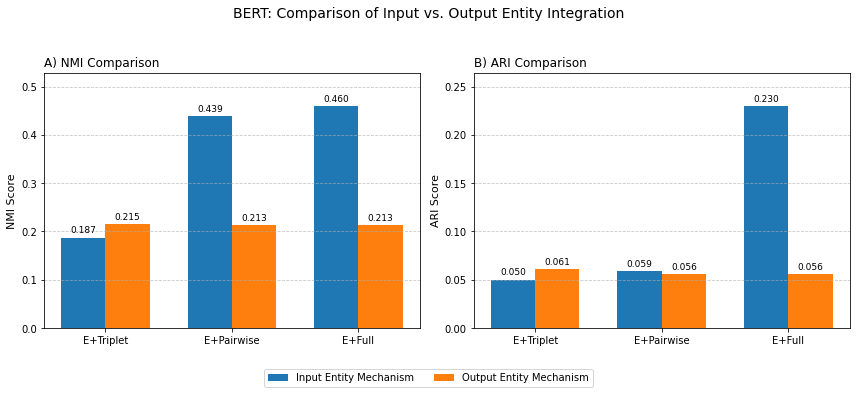} 
    \caption{Comparison of NMI (A) and ARI (B) for BERT using Input vs. Output-level entity integration across contrastive loss types.}
    \label{fig:bert_entity_comparison}
    \vspace{-10pt}
\end{figure}

The results indicate that incorporating this specific output-level entity-aware loss did not improve clustering performance and frequently led to substantial degradation compared to the standard contrastive objectives alone. For instance, comparing the standard Pairwise Loss (\(\mathcal{L}_p\) only) to the configuration adding the entity loss (\(\mathcal{L}_p + w_e \mathcal{L}_e\)), RoBERTa's NMI dropped significantly from 0.702 to 0.120, and DarkBERT's NMI decreased from 0.592 to 0.339 (detailed results omitted for brevity). This suggests that directly forcing alignment between overall post embeddings and specific entity embeddings via this auxiliary loss interfered with learning representations optimal for the primary task of clustering posts based on the overall event context. Figure~\ref{fig:bert_entity_comparison} visually compares the impact of the input-level versus output-level entity-aware mechanisms specifically for the BERT model across the different loss configurations. As shown by the orange bars (Output Entity Mechanism), performance (both NMI and ARI) is consistently low and generally worse than the corresponding results for the Input Entity Mechanism (blue bars), which itself showed mixed results (Finding 4). The output-level entity-aware mechanism, in particular, struggled to surpass even baseline levels achieved by standard contrastive fine-tuning. Consistent with Finding 4, the specific entity-aware mechanisms explored here, whether at the input or output level, proved less effective overall than standard contrastive fine-tuning for this task.

\textbf{Overall Performance Summary.}
Fine-tuned Transformer embeddings effectively cluster security events from forums. The best performance came from standard contrastive fine-tuning on domain-adapted models: DarkBERT (Pairwise Loss) achieved the highest NMI (0.712), and CySecBERT (Pairwise Loss) the highest ARI (0.377). While an input-level entity-aware mechanism significantly improved BERT (Combined Loss) and yielded competitive results for DarkBERT (higher ARI), it wasn't universally better across all tested models and configurations. An output-level entity-aware loss proved ineffective. These results highlight the importance of domain adaptation and contrastive objectives, while suggesting the input-entity approach has model-specific potential that could be optimized in future work.

\subsection{Ranking Performance}
\label{subsec:ranking_performance}

A key objective of \model{} is to move beyond simple event detection towards actionable intelligence by prioritizing the discovered event clusters based on their potential operational significance. This addresses the critical challenge of analyst overload when faced with numerous alerts generated from noisy data sources like hacker forums~\cite{sans2023ctisurvey}. We evaluate the prioritization mechanism by analyzing the characteristics of events ranked at different positions using qualitative case studies and by comparing the overall ranking order produced by \model{} against simpler heuristics.

For this evaluation, we applied the \model{} ranker using a default weighting scheme (\( w_{T}=0.35, w_{V}=0.25, w_{R}=0.20, w_{C}=0.20 \)) designed to prioritize event timeliness, reflecting the operational need to surface recent and active threats, while giving balanced consideration to other factors. The relevance component was disabled (\( V=0 \)) to assess the general, query-agnostic ranking performance. The reference time was set to the latest post timestamp in the dataset, and the ranker was applied directly to the 70 curated ground truth event clusters. This allowed us to assess how the prioritization score behaves across a set of known, distinct events.

\subsubsection{Qualitative Analysis: Case Studies}
\label{subsubsec:ranking_case_studies}

To qualitatively assess the ranking's ability to surface operationally relevant events, we examined specific event clusters ranked at different positions (top, middle, bottom) by \model{}. Table~\ref{tab:ranking_case_studies} presents the calculated priority scores (Overall Score, Timeliness T, Credibility R, Completeness C) for three representative examples from our curated ground truth set.

\begin{table}[tbp]
\centering

\caption{Ranking Scores for Case Study Events.}
\label{tab:ranking_case_studies}

\begin{tabular}{@{} l l c c c c @{}}
\toprule
\textbf{Rank} & \textbf{Event (Approx. Date)} & \textbf{Score} & \textbf{T} & \textbf{R} & \textbf{C} \\
\midrule
1 & Optus Breach (Sep/Oct 22) & 0.414 & 0.630 & 0.584 & 0.382 \\
\midrule
19 & ThaiTradeFair (Dec 22) & 0.202 & 0.204 & 0.275 & 0.379 \\
\midrule
37 & LockBit 3.0 (Jun-Dec 22) & 0.090 & 0.010 & 0.350 & 0.083 \\
\bottomrule
\end{tabular}
\vspace{-10pt}
\end{table}

\textbf{Rank 1: Optus Data Breach (Sep/Oct 2022).} This high-profile event, associated with 52 posts in our dataset, achieved the top rank (Score: 0.414). As shown in Table~\ref{tab:ranking_case_studies}, its ranking is primarily driven by an excellent Timeliness score (T=0.630), reflecting recent and sustained activity, combined with strong Credibility (R=0.584) based on contributing authors. Its Completeness (C=0.382) was moderate. This outcome aligns with the goal of prioritizing major, ongoing incidents involving reputable forum members.

\textbf{Rank 19: ThaiTradeFair.com Leak (Dec 2022).} Representing a mid-tier rank (Score: 0.202), this less prominent data leak discussion comprised only 6 posts. Its lower Timeliness (T=0.204) and significantly lower author Credibility score (R=0.275) contributed to its moderate ranking. The event showed reasonable Completeness (C=0.379), indicating some relevant entities were mentioned despite the low post count.

\textbf{Rank 37: LockBit 3.0 Infrastructure Discussion (June-Dec 2022).} this cluster discussed older LockBit infrastructure and policies across 7 posts. Despite moderate author Credibility (R=0.350), its very low score stems directly from extremely poor Timeliness (T=0.010) and low Completeness (C=0.083). This correctly deprioritizes older, less detailed discussions relative to more current or richer events.

These case studies demonstrate that the Priority Score integrates multiple dimensions to produce a nuanced ranking. High ranks are not solely determined by activity (post count) or recency alone, but by a combination reflecting potential significance based on timeliness, credibility, and information richness. The system effectively differentiates between major ongoing events, smaller or less credible incidents, and older, less complete discussions based on the calculated metrics.

\subsubsection{Comparison with Baseline Ranking Orders}
\label{subsubsec:ranking_baselines}

We also compared the ranking order produced by \model{} against two simpler baseline heuristics:
\begin{itemize}
\item \textbf{Recency Rank}: Events ranked solely by the timestamp of their latest post (most recent first).
\item \textbf{Activity Rank}: Events ranked solely by the total number of posts within their cluster (highest count first).
\end{itemize}

Observing the top-ranked events for each method (Table~\ref{tab:top_ranked_comparison}) highlights the different perspectives provided. \model{}'s top rank (Optus) is also top-ranked by Recency but only third by Activity. HSYS, ranked second by \model{} and first by Activity (due to its high post count of 116), is ranked second by Recency. The DDoS tool discussion, third in \model{}'s ranking, appears third in Recency but only eighth in Activity. This illustrates that \model{}'s multi-dimensional score produces a distinct ordering that balances factors beyond simple recency or volume, potentially offering a more holistic view of event significance compared to unidimensional heuristics.

\begin{table}[tbp] 
\centering
\caption{Comparison of Top 3 Ranked Events by Different Methods}
\label{tab:top_ranked_comparison}
\resizebox{\columnwidth}{!}{
\begin{tabular}{llll}
\toprule
\textbf{Rank} & \textbf{\model{}} & \textbf{Recency Rank} & \textbf{Activity Rank} \\
\midrule
1 & Optus Data Breach & Optus Data Breach & HSYS Leak \\
2 & HSYS Leak & HSYS Leak & Banorte Leak \\
3 & DDoS Tool Method & DDoS Tool Method & Optus Data Breach \\
\bottomrule
\end{tabular}
} 
\vspace{-10pt} 
\end{table}

\textbf{Summary.} The evaluation demonstrates \model{}'s event prioritization mechanism generates a plausible ranking based on Timeliness, Credibility, and Completeness. Qualitative case studies confirm the interpretability of the ranking and its ability to differentiate events based on these factors. Comparison with baseline methods (~\ref{subsubsec:ranking_baselines}) shows that \model{} produces a distinct event ordering by integrating multiple dimensions, offering a potentially more comprehensive assessment of event significance than rankings based solely on recency or activity. The presented results highlight the functionality and potential utility of the proposed ranking approach for CTI analysts navigating high-volume forum data.
\section{Discussion}
\label{sec:discussion}

Our results demonstrate that \model{} can effectively identify and cluster security events from noisy forum data. The strong performance of domain-specific Transformers fine-tuned with contrastive objectives highlights the importance of contextual representation for this task. The framework's ability to aggregate fragmented discussions into coherent, prioritized events offers a significant step towards managing the deluge of data from underground forums. However, the work has several limitations that open important avenues for future research.

\subsection{Limitations}
\label{subsec:limitations}

A primary limitation is the static nature of our current evaluation. The clustering is performed on a snapshot of the data and does not explicitly model the temporal evolution of events, which is critical for tracking threats as they unfold. Second, our exploration of entity-aware mechanisms, while showing model-specific promise (Finding 4), did not yield a universally superior integration strategy. A deeper investigation into how entity information can robustly enhance event clustering is still needed.

Third, the Priority Score's ranking is sensitive to the configured weights of its components (Timeliness, Credibility, etc.). Our evaluation used a default set of weights, but a comprehensive sensitivity analysis is required to understand how different operational priorities would alter the final event ordering. Finally, our metrics for credibility and timeliness are derived solely from internal forum metadata. They do not incorporate external ground truth, which could provide a more objective measure of an event's real-world impact and the timeliness of its detection.

\subsection{Future Work and Broader Impact}
\label{subsec:future_work}

These limitations highlight several promising research directions. Addressing the static nature of the evaluation, future work should focus on incorporating timestamps and the temporal relationships between posts to enable dynamic event tracking and more refined clustering. This would allow the system to model the entire lifecycle of a security event, from its inception to its resolution.

Regarding the prioritization mechanism, we acknowledge that our weighted linear combination is a baseline approach. Future research could explore more advanced multi-objective ranking techniques, such as skyline queries, to provide analysts with a more nuanced set of non-dominated ``best'' events across different dimensions, rather than a single linear ranking. Furthermore, the ranking could be validated and enriched by correlating its outputs with external data sources. For example, the framework's timeliness could be benchmarked against public threat intelligence reports to quantify the early-warning value of forum monitoring, and event completeness could be compared against official CVSS scores.

The framework also provides a foundation for identifying significant actors, a suggestion that requires further validation. By calculating an actor's aggregated priority score based on their participation in high-ranking events ($\text{ActorScore}(a) = \sum_{c_i \in \mathcal{C}_a} s_i$), one could develop an event-centric measure of user influence. Validating this data-driven approach against known threat actor profiles is a key next step.

Finally, scaling the framework for real-time, high-volume data ingestion and conducting empirical evaluations of its effectiveness in a live operational setting are essential for transitioning \model{} from a research prototype to a practical tool for security analysts. Such a deployment would also allow for a direct assessment of how the system helps mitigate analyst workload and improves response times to emerging threats.

\subsection{Ethical Considerations}
\label{subsec:ethical_consideration}

Our research uses the CrimeBB academic dataset~\cite{Pastrana2018CrimeBB}, sourced from public forums under a Data Use Agreement (DUA), which prohibits sharing the raw data. This data is sensitive and was handled ethically: analysis used only public text and involved no user interaction. To support reproducibility despite data restrictions, the source code for the \model{} framework covering embedding model training, clustering, and prioritization is available at \url{https://github.com/yasirech-chammakhy/EventHunter}.

\section{Conclusion}
\label{sec:conclusion}

This paper introduced \model{}, an unsupervised framework designed to automatically detect and prioritize security events within noisy, fragmented hacker forum discussions. Our approach addresses fundamental challenges in extracting actionable intelligence by integrating entity-aware contrastive embeddings tailored for security semantics, robust density-based clustering for aggregating related posts into events, and a systematic prioritization mechanism based on CTI quality metrics. \model{} demonstrates a scalable methodology for moving from raw forum data to prioritized, operationally relevant threat intelligence, enabling security analysts to focus effectively on the most significant emerging risks discussed within the cybercrime ecosystem.

\section*{Acknowledgment}
This work was supported by Deloitte Morocco Cyber Center

\bibliographystyle{plain}
\bibliography{references}

\begin{thebibliography}{10}

\bibitem{aghaei2022securebert}
Ehsan Aghaei, Xi~Niu, Waseem Shadid, and Ehab Al-Shaer.
\newblock Securebert: A domain-specific language model for cybersecurity.
\newblock {\em arXiv}, October 2022.

\bibitem{alam2022cyner}
Md~Tanvirul Alam, Dipkamal Bhusal, Youngja Park, and Nidhi Rastogi.
\newblock Cyner: A python library for cybersecurity named entity recognition, 2022.
\newblock Accessed: 2025-03-15.

\bibitem{Alkhodair2020Detecting}
Sarah~A. Alkhodair, Steven H.~H. Ding, Benjamin C.~M. Fung, and Junqiang Liu.
\newblock Detecting breaking news rumors of emerging topics in social media.
\newblock {\em Information Processing \& Management}, 57(2):102018, March 2020.

\bibitem{amadou2024eurekha}
Abdoul Nasser~Hassane Amadou, Anas Motii, Saida Elouardi, and El~Houcine Bergou.
\newblock {EUREKHA: Enhancing User Representation for Key Hackers Identification in Underground Forums}.
\newblock In {\em 2024 IEEE 23rd International Conference on Trust, Security and Privacy in Computing and Communications (TrustCom)}, pages 387--398. IEEE, 2024.

\bibitem{amadou2024hc}
Abdoul Nasser~Hassane Amadou, Anas Motii, and Mohammed Jouhari.
\newblock {HC-HackerRank: Identifying Key Hackers in Cybercrime Social Network Forums}.
\newblock In {\em 2024 7th International Conference on Advanced Communication Technologies and Networking (CommNet)}, pages 1--8. IEEE, 2024.

\bibitem{arazzi2023nlp}
Marco Arazzi, Dincy~R. Arikkat, Serena Nicolazzo, Antonino Nocera, Rafidha Rehiman K.~A., P.~Vinod, and Mauro Conti.
\newblock {NLP-Based Techniques for Cyber Threat Intelligence}.
\newblock {\em arXiv preprint}, arXiv:2311.08807, November 15 2023.

\bibitem{Ashktorab2019Using}
Zahra Ashktorab, Christopher Brown, Manojit Nandi, and Aron Culotta.
\newblock Using twitter data to monitor natural disaster social dynamics: A recurrent neural network approach with word embeddings and kernel density estimation.
\newblock {\em Sensors}, 19(7):1746, 2019.

\bibitem{ashok2023promptner}
Dhananjay Ashok and Zachary~C. Lipton.
\newblock {PromptNER: Prompting For Named Entity Recognition}.
\newblock {\em arXiv}, June 2023.
\newblock arXiv preprint arXiv:2305.15444.

\bibitem{bose2019novel}
Avishek Bose, Vahid Behzadan, Carlos Aguirre, and William~H. Hsu.
\newblock A novel approach for detection and ranking of trendy and emerging cyber threat events in twitter streams.
\newblock In {\em Proceedings of the 2019 IEEE/ACM International Conference on Advances in Social Networks Analysis and Mining}, pages 871--878. ACM, 2019.

\bibitem{campello2013density}
Ricardo~JGB Campello, Davoud Moulavi, and Joerg Sander.
\newblock Density-based clustering based on hierarchical density estimates.
\newblock In {\em Advances in Knowledge Discovery and Data Mining}, volume 7819, pages 160--172. Springer, 2013.

\bibitem{chen2024stix}
Sheng-Shan Chen, Ren-Hung Hwang, Asad Ali, Ying-Dar Lin, Yu-Chih Wei, and Tun-Wen Pai.
\newblock Improving quality of indicators of compromise using stix graphs.
\newblock {\em Computers \& Security}, 144:103972, September 2024.

\bibitem{chen2023enhancing}
Sheng-Shan Chen, Ren-Hung Hwang, Chin-Yu Sun, Ying-Dar Lin, and Tun-Wen Pai.
\newblock Enhancing cyber threat intelligence with named entity recognition using bert-crf.
\newblock In {\em GLOBECOM 2023 - 2023 IEEE Global Communications Conference}, pages 7532--7537, Kuala Lumpur, Malaysia, 2023. IEEE.

\bibitem{chen2024survey}
Yiren Chen, Mengjiao Cui, Ding Wang, Yiyang Cao, Peian Yang, Bo~Jiang, Zhigang Lu, and Baoxu Liu.
\newblock A survey of large language models for cyber threat detection.
\newblock {\em Computers \& Security}, 145:104016, October 2024.

\bibitem{cheng2024novel}
Qi~Cheng, Liqiong Chen, Zhixing Hu, Juan Tang, Qiang Xu, and Binbin Ning.
\newblock {A Novel Prompting Method for Few-Shot NER via LLMs}.
\newblock {\em Natural Language Processing Journal}, 8:100099, September 2024.

\bibitem{cui2024tweezers}
Jian Cui, Hanna Kim, Eugene Jang, Dayeon Yim, Kicheol Kim, Yongjae Lee, Jin-Woo Chung, Seungwon Shin, and Xiaojing Liao.
\newblock Tweezers: A framework for security event detection via event attribution-centric tweet embedding.
\newblock arXiv preprint arXiv:2409.08221, 2024.

\bibitem{Cybersixgill2024}
Cybersixgill.
\newblock Cyber threat intelligence survey, 2024.
\newblock Accessed: 2025-02-27.

\bibitem{deliu2018collecting}
Isuf Deliu, Carl Leichter, and Katrin Franke.
\newblock Collecting cyber threat intelligence from hacker forums via a two-stage, hybrid process using support vector machines and latent dirichlet allocation.
\newblock In {\em 2018 IEEE International Conference on Big Data (Big Data)}, pages 5008--5013, Seattle, WA, USA, 2018. IEEE.

\bibitem{devlin2019bert}
Jacob Devlin et~al.
\newblock Bert: Pre-training of deep bidirectional transformers for language understanding.
\newblock In {\em NAACL}, 2019.

\bibitem{dong2018new}
Fangzhou Dong, Shaoxian Yuan, and Liang Liu.
\newblock New cyber threat discovery from darknet marketplaces.
\newblock In {\em 2018 IEEE International Conference on Big Data and Artificial Intelligence (ICBDAI)}, pages 69--72. IEEE, 2018.

\bibitem{Elouardi2024HybridCNNLLM}
Saida Elouardi, Anas Motii, Mohammed Jouhari, Abdoul Amadou, and Mustapha Hedabou.
\newblock A survey on hybrid-cnn and llms for intrusion detection systems: Recent iot datasets.
\newblock {\em IEEE Access}, PP:1--1, 2024.
\newblock Published: November 26, 2024.

\bibitem{fang2020detecting}
Yong Fang, Jian Gao, Zhonglin Liu, and Cheng Huang.
\newblock Detecting cyber threat event from twitter using idcnn and bilstm.
\newblock {\em Applied Sciences}, 10(17):5922, 2020.

\bibitem{fang2019analyzing}
Yong Fang, Yusong Guo, Cheng Huang, and Liang Liu.
\newblock Analyzing and identifying data breaches in underground forums.
\newblock {\em IEEE Access}, 7:48770--48777, 2019.

\bibitem{Fedoryszak2019RealTime}
Mateusz Fedoryszak, Brent Frederick, Vijay Rajaram, and Changtao Zhong.
\newblock Real-time event detection on social data streams.
\newblock In {\em Proceedings of the 25th ACM SIGKDD International Conference on Knowledge Discovery \& Data Mining}, pages 2774--2782, New York, NY, USA, 2019. Association for Computing Machinery.

\bibitem{fiorini2023cysecbert}
Nicolas Fiorini et~al.
\newblock Cysecbert: Cybersecurity language model.
\newblock In {\em IEEE International Conference on Big Data}, 2023.

\bibitem{gao2021simcse}
Tianyu Gao, Xingcheng Yao, and Danqi Chen.
\newblock Simcse: Simple contrastive learning of sentence embeddings.
\newblock {\em arXiv preprint arXiv:2104.08821}, 2021.

\bibitem{geras2024quality}
Thomas Geras and Thomas Schreck.
\newblock The ‘big beast to tackle’: Practices in quality assurance for cyber threat intelligence.
\newblock In {\em The 27th International Symposium on Research in Attacks, Intrusions and Defenses}, pages 337--352, Padua, Italy, 2024. ACM.

\bibitem{Hagras2017Towards}
Mohamed Hagras, Ghada Hassan, and Nadine Farag.
\newblock Towards natural disasters detection from twitter using topic modelling.
\newblock In {\em 2017 European Conference on Electrical Engineering and Computer Science (EECS)}, pages 272--279. IEEE, 2017.

\bibitem{hubert1985comparing}
Lawrence Hubert and Phipps Arabie.
\newblock Comparing partitions.
\newblock {\em Journal of Classification}, 2(1):193--218, 1985.

\bibitem{sans2023ctisurvey}
SANS Institute.
\newblock 2023 cti survey: Keeping up with a changing threat landscape, 2023.

\bibitem{jin2023darkbert}
Seungjun Jin et~al.
\newblock Darkbert: A language model for the dark side of the internet.
\newblock In {\em ACL}, 2023.

\bibitem{kadoguchi2019exploring}
Masashi Kadoguchi, Shota Hayashi, Masaki Hashimoto, and Akira Otsuka.
\newblock Exploring the dark web for cyber threat intelligence using machine learning.
\newblock In {\em 2019 IEEE International Conference on Intelligence and Security Informatics (ISI)}, pages 200--202. IEEE, 2019.

\bibitem{khandpur2017crowdsourcing}
Rupinder~Paul Khandpur, Taoran Ji, Steve Jan, Gang Wang, Chang-Tien Lu, and Naren Ramakrishnan.
\newblock Crowdsourcing cybersecurity: Cyber attack detection using social media.
\newblock \url{https://doi.org/10.48550/arXiv.1702.07745}, 2017.
\newblock arXiv:1702.07745.

\bibitem{komecoglu2024event}
Basak Kömeçoglu and Burcu Yilmaz.
\newblock Event graph-based news clustering: The role of named entity-centered subgraphs.
\newblock {\em IEEE Access}, PP:1--1, January 2024.

\bibitem{li2021nedetector}
Ying Li, Jiaxing Cheng, Cheng Huang, Zhouguo Chen, and Weina Niu.
\newblock Nedetector: Automatically extracting cybersecurity neologisms from hacker forums.
\newblock {\em Journal of Information Security and Applications}, 58:102784, May 2021.

\bibitem{liu2019roberta}
Yinhan Liu et~al.
\newblock Roberta: A robustly optimized bert pretraining approach.
\newblock {\em arXiv preprint arXiv:1907.11692}, 2019.

\bibitem{logeswaran2019zero}
Lajanugen Logeswaran, Ming-Wei Chang, Kenton Lee, Kristina Toutanova, Jacob Devlin, and Honglak Lee.
\newblock Zero-shot entity linking by reading entity descriptions.
\newblock In Anna Korhonen, David Traum, and Llu\'is M\`arquez, editors, {\em Proceedings of the 57th Annual Meeting of the Association for Computational Linguistics}, pages 3449--3460, Florence, Italy, 2019. Association for Computational Linguistics.

\bibitem{loshchilov2019decoupled}
Ilya Loshchilov and Frank Hutter.
\newblock Decoupled weight decay regularization.
\newblock {\em arXiv preprint arXiv:1711.05101}, 2019.

\bibitem{CyberMagazine2024}
Cyber Magazine.
\newblock The rapidly evolving threat landscape of 2024, 2024.
\newblock Accessed: 2025-02-27.

\bibitem{manatova2024understand}
Dalyapraz Manatova, Charles DeVries, and Sagar Samtani.
\newblock Understand your shady neighborhood: An approach for detecting and investigating hacker communities.
\newblock {\em Decision Support Systems}, 184:114271, September 2024.

\bibitem{moreno2023cream}
Felipe Moreno-Vera, Mateus Nogueira, Cainã Figueiredo, Daniel~Sadoc Menasché, Miguel Bicudo, Ashton Woiwood, Enrico Lovat, Anton Kocheturov, and Leandro~Pfleger de~Aguiar.
\newblock Cream skimming the underground: Identifying relevant information points from online forums.
\newblock arXiv preprint arXiv:2308.02581, 2023.
\newblock Accessed: 2025-03-28.

\bibitem{mouiche2025entity}
Inoussa Mouiche and Sherif Saad.
\newblock Entity and relation extractions for threat intelligence knowledge graphs.
\newblock {\em Computers \& Security}, 148:104120, January 2025.

\bibitem{paladini2024hackerforums}
Tommaso Paladini, Lara Ferro, Mario Polino, Stefano Zanero, and Michele Carminati.
\newblock You might have known it earlier: Analyzing the role of underground forums in threat intelligence.
\newblock In {\em Proceedings of the 27th International Symposium on Research in Attacks, Intrusions and Defenses (RAID)}. ACM, 2024.

\bibitem{Parekh2024Event}
Tanmay Parekh, Anh Mac, Jiarui Yu, Yuxuan Dong, Syed Shahriar, Bonnie Liu, Eric Yang, et~al.
\newblock Event detection from social media for epidemic prediction.
\newblock In Kevin Duh, Helena Gomez, and Steven Bethard, editors, {\em Proceedings of the 2024 Conference of the North American Chapter of the Association for Computational Linguistics: Human Language Technologies (Volume 1: Long Papers)}, pages 5758--5783, Mexico City, Mexico, 2024. Association for Computational Linguistics.

\bibitem{Pastrana2018CrimeBB}
Sergio Pastrana, Daniel~R. Thomas, Alice Hutchings, and Richard Clayton.
\newblock {CrimeBB: Enabling Cybercrime Research on Underground Forums at Scale}.
\newblock In {\em Proceedings of the 2018 World Wide Web Conference on World Wide Web - WWW '18}, pages 1845--1854. ACM Press, 2018.

\bibitem{paszke2019pytorch}
Adam Paszke, Sam Gross, Francisco Massa, Adam Lerer, James Bradbury, Gregory Chanan, Trevor Killeen, Zeming Lin, Natalia Gimelshein, Luca Antiga, et~al.
\newblock Pytorch: An imperative style, high-performance deep learning library.
\newblock In {\em Advances in Neural Information Processing Systems}, volume~32. Curran Associates, Inc., 2019.

\bibitem{pedregosa2011scikit}
Fabian Pedregosa, Gael Varoquaux, Alexandre Gramfort, Vincent Michel, Bertrand Thirion, Olivier Grisel, Mathieu Blondel, Peter Prettenhofer, Ron Weiss, Vincent Dubourg, et~al.
\newblock Scikit-learn: Machine learning in python.
\newblock {\em Journal of Machine Learning Research}, 12:2825--2830, 2011.

\bibitem{rahman2023what}
Md~Rayhanur Rahman, Rezvan Mahdavi-Hezaveh, and Laurie Williams.
\newblock What are the attackers doing now? automating cyber threat intelligence extraction from text on pace with the changing threat landscape: A survey.
\newblock {\em ACM Computing Surveys}, 55(12):1--36, 2023.

\bibitem{sadlek2025severity}
Lukáš Sadlek, Muhammad~Mudassar Yamin, Pavel Čeleda, and Basel Katt.
\newblock Severity-based triage of cybersecurity incidents using kill chain attack graphs.
\newblock {\em Journal of Information Security and Applications}, 89:103956, March 2025.

\bibitem{samtani2017exploring}
Sagar Samtani, Ryan Chinn, Hsinchun Chen, and Jay Nunamaker.
\newblock Exploring emerging hacker assets and key hackers for proactive cyber threat intelligence.
\newblock {\em Journal of Management Information Systems}, 34(4):1023--1053, 2017.

\bibitem{sapienza2017early}
Anna Sapienza, Alessandro Bessi, Saranya Damodaran, Paulo Shakarian, Kristina Lerman, and Emilio Ferrara.
\newblock Early warnings of cyber threats in online discussions.
\newblock In {\em 2017 IEEE International Conference on Data Mining Workshops (ICDMW)}, pages 667--674, New Orleans, LA, 2017. IEEE.

\bibitem{saravanakumar2021event}
Kailash~Karthik Saravanakumar, Miguel Ballesteros, Muthu~Kumar Chandrasekaran, and Kathleen McKeown.
\newblock Event-driven news stream clustering using entity-aware contextual embeddings.
\newblock In Paola Merlo, Jörg Tiedemann, and Reut Tsarfaty, editors, {\em Proceedings of the 16th Conference of the European Chapter of the Association for Computational Linguistics: Main Volume}, pages 2330--2340, Online, 2021. Association for Computational Linguistics.

\bibitem{sceller2017sonar}
Quentin Sceller, Elmouatez Karbab, Mourad Debbabi, and Farkhund Iqbal.
\newblock {SONAR: Automatic Detection of Cyber Security Events over the Twitter Stream}.
\newblock In {\em Proceedings of the 2017 ACM on Conference on Information and Knowledge Management (CIKM)}, 2017.

\bibitem{shin2020cybersecurity}
Hyejin Shin, WooChul Shim, Jiin Moon, Jae Seo, Sol Lee, and Yong Hwang.
\newblock Cybersecurity event detection with new and re-emerging words.
\newblock In {\em Proceedings of the 2020 ACM SIGKDD Workshop on Cybersecurity and Intelligence}. ACM, 2020.

\bibitem{staykovski2019dense}
Todor Staykovski, Alberto Barrón-Cedeño, Giovanni Da~San~Martino, and Preslav Nakov.
\newblock Dense vs. sparse representations for news stream clustering.
\newblock In {\em Proceedings of the 2nd International Workshop on Narrative Extraction from Texts (Text2Story@ECIR)}, pages 43--48, Cologne, Germany, 2019.
\newblock Workshop held at ECIR 2019.

\bibitem{tang2023trigger}
Mengmeng Tang, Yuanbo Guo, Qingchun Bai, and Han Zhang.
\newblock Trigger-free cybersecurity event detection based on contrastive learning.
\newblock {\em The Journal of Supercomputing}, 79:20984--21007, 2023.

\bibitem{tundis2022feature}
Andrea Tundis, Samuel Ruppert, and Max Mühlhäuser.
\newblock A feature-driven method for automating the assessment of osint cyber threat sources.
\newblock {\em Computers \& Security}, 113:102576, 2022.

\bibitem{vinh2010information}
Nguyen~Xuan Vinh, Julien Epps, and James Bailey.
\newblock Information theoretic measures for clusterings comparison: Variants, properties, normalization and correction for chance.
\newblock In {\em Journal of Machine Learning Research}, volume~11, pages 2837--2854, 2010.

\bibitem{wang2022aptner}
Xuren Wang, Songheng He, Zihan Xiong, Xinxin Wei, Zhengwei Jiang, Sihan Chen, and Jun Jiang.
\newblock Aptner: A specific dataset for ner missions in cyber threat intelligence field.
\newblock In {\em 2022 IEEE 25th International Conference on Computer Supported Cooperative Work in Design (CSCWD)}, pages 1233--1238, Hangzhou, China, 2022. IEEE.

\bibitem{wang2020dnrti}
Xuren Wang, Xinpei Liu, Shengqin Ao, Ning Li, Zhengwei Jiang, Zongyi Xu, Zihan Xiong, Xiong Mengbo, and Xiaoqing Zhang.
\newblock Dnrti: A large-scale dataset for named entity recognition in threat intelligence.
\newblock In {\em 2020 IEEE 19th International Conference on Trust, Security and Privacy in Computing and Communications (TrustCom)}, pages 1632--1639. IEEE, 2020.

\bibitem{yagcioglu2019detecting}
Semih Yagcioglu, Mehmet~Saygin Seyfioglu, Begum Citamak, Batuhan Bardak, Seren Guldamlasioglu, Azmi Yuksel, and Emin~Islam Tatli.
\newblock Detecting cybersecurity events from noisy short text.
\newblock pages 1366--1372, 2019.

\bibitem{Yousefinaghani2019Assessment}
Samira Yousefinaghani, Rozita Dara, Zvonimir Poljak, Theresa Bernardo, and Shayan Sharif.
\newblock The assessment of twitter’s potential for outbreak detection: Avian influenza case study.
\newblock {\em Scientific Reports}, 9, 2019.

\bibitem{zhang2018idetector}
Yiming Zhang, Yujie Fan, Shifu Hou, Jian Liu, Yanfang Ye, and Thirimachos Bourlai.
\newblock {iDetector}: Automate underground forum analysis based on heterogeneous information network.
\newblock In {\em 2018 IEEE/ACM International Conference on Advances in Social Networks Analysis and Mining (ASONAM)}, pages 1071--1078, Barcelona, 2018. IEEE.

\bibitem{zibak2022threat}
Adam Zibak, Clemens Sauerwein, and Andrew~C. Simpson.
\newblock Threat intelligence quality dimensions for research and practice.
\newblock {\em Digital Threats: Research and Practice}, 3(4):1--22, 2022.

\end{thebibliography}

\end{document}